\documentclass{article}

\usepackage{PRIMEarxiv}

\usepackage[utf8]{inputenc} % allow utf-8 input
\usepackage[T1]{fontenc}    % use 8-bit T1 fonts
\usepackage{hyperref}       % hyperlinks
\usepackage{url}            % simple URL typesetting
\usepackage{booktabs}       % professional-quality tables
\usepackage{amsfonts}       % blackboard math symbols
\usepackage{nicefrac}       % compact symbols for 1/2, etc.
\usepackage{microtype}      % microtypography
\usepackage{lipsum}
\usepackage{fancyhdr}       % header
\usepackage{graphicx}       % graphics
\usepackage{multirow}
\usepackage{multicol}
\usepackage{etoolbox}
\usepackage{subcaption}
\usepackage{amsmath}
\usepackage{placeins}
\graphicspath{{media/}}     % organize your images and other figures under media/ folder
\title{High-fidelity simulation of pebble beds:  Toward an improved understanding of the wall channeling effect
\thanks{\textit{\underline{Citation}}: 
\textbf{Authors. Title. Pages.... DOI:000000/11111.}} 
}

\author{
  David Reger, Elia Merzari, and Saya Lee \\
  Pennsylvania State University \\
  University Park, PA 16803\\
  \texttt{dzr5281@psu.edu ebm5351@psu.edu} \\
   \And
  Paolo Balestra \\
  Idaho National Laboratory \\
  Idaho Falls, ID 83415\\
  \texttt{paolo.balestra@inl.gov} \\
  \And
  Yassin Hassan \\
  Texas A\&M University  \\
  College Station, TX 77843 \\
  \texttt{y-hassan@tamu.edu}
  \And
}

\begin{document}
\maketitle

\begin{abstract}
Wall channeling is a phenomena of interest for Pebble Bed Reactors (PBRs) where flow is diverted into high-porosity regions near the wall. This diversion of flow can have a significant impact on maximum fuel temperatures and core bypass flow. Porous media models that are currently used to model PBRs for design scoping and transient simulation are lacking in their capabilities to model the wall channel effect. Recent efforts at Penn State have produced an improved porous media pressure drop equation that is more capable of modeling the velocity variations caused by the wall channel effect in a porous media model. Several pebble beds were divided into concentric rings of $0.05D_{peb}$, and average flow quantities and porosities were extracted for the ring. A correlation between the form loss coefficient and the local ring porosity was found, allowing for the addition of a correction factor to the form loss term of the KTA equation. The developed correlation was purely empirical, and thus a more thorough understanding of the underlying flow phenomena is desired. This study investigates geometric and flow features that can explain the observed correlation between the form coefficient and the local porosity that was used to generate the improved pressure drop equation. The solid surface area to volume ratio $S_v$ along with the production of Turbulent Kinetic Energy (TKE) is analyzed. A relationship between $S_v$ and the local porosity and an inverse relationship between the negative TKE production and the local porosity were found, pointing to the idea that inertial effects caused by different pore geometry in each ring contribute to the variation of the form constant with the local porosity.
\end{abstract}

% keywords can be removed

\section{Introduction}
The Pebble Bed Reactor (PBR) design has seen a resurgence in interest in recent years. PBRs are currently being developed in the United States and China is currently constructing several PBRs. As these systems are approaching more widespread deployment, fast and accurate simulation tools are necessary for design scoping and simulation of accident scenarios.

Porous media modeling is the current state-of-the-art for intermediate-fidelity simulation of packed beds. Especially in the case of randomly organized porous media like pebble beds, resolution of the complex void regions and all fluid-solid interfaces is incredibly expensive, requiring billions of gridpoints \cite{Wu2010}. It is worth noting, however, that these pebble-resolved computations are not impossible, as recent increases in computing power have made such calculations feasible \cite{Cardinal}. Regardless, the immense cost involved with pebble-resolved simulations makes them unrealistically expensive for design scoping and plant-level simulation. Porous media models mend this issue by homogenizing the porous media with spatial averaging, reducing the computational cost by several orders of magnitude.
Closure models are also required to approximate the effects of small flow features on the macroscale behavior of the flow. These closure models provide estimations of drag coefficients, effective conductivities, and interphase heat transfer coefficients among other parameters \cite{Pronghorn2021}. The accuracy of the closure models will directly influence the accuracy of the porous media model as a whole, and thus it is important to ensure that the models effectively represent real-world physics.

\begin{figure}
    \centering
    \begin{subfigure}[]{0.48\textwidth}
       \includegraphics[width=1.0\textwidth]{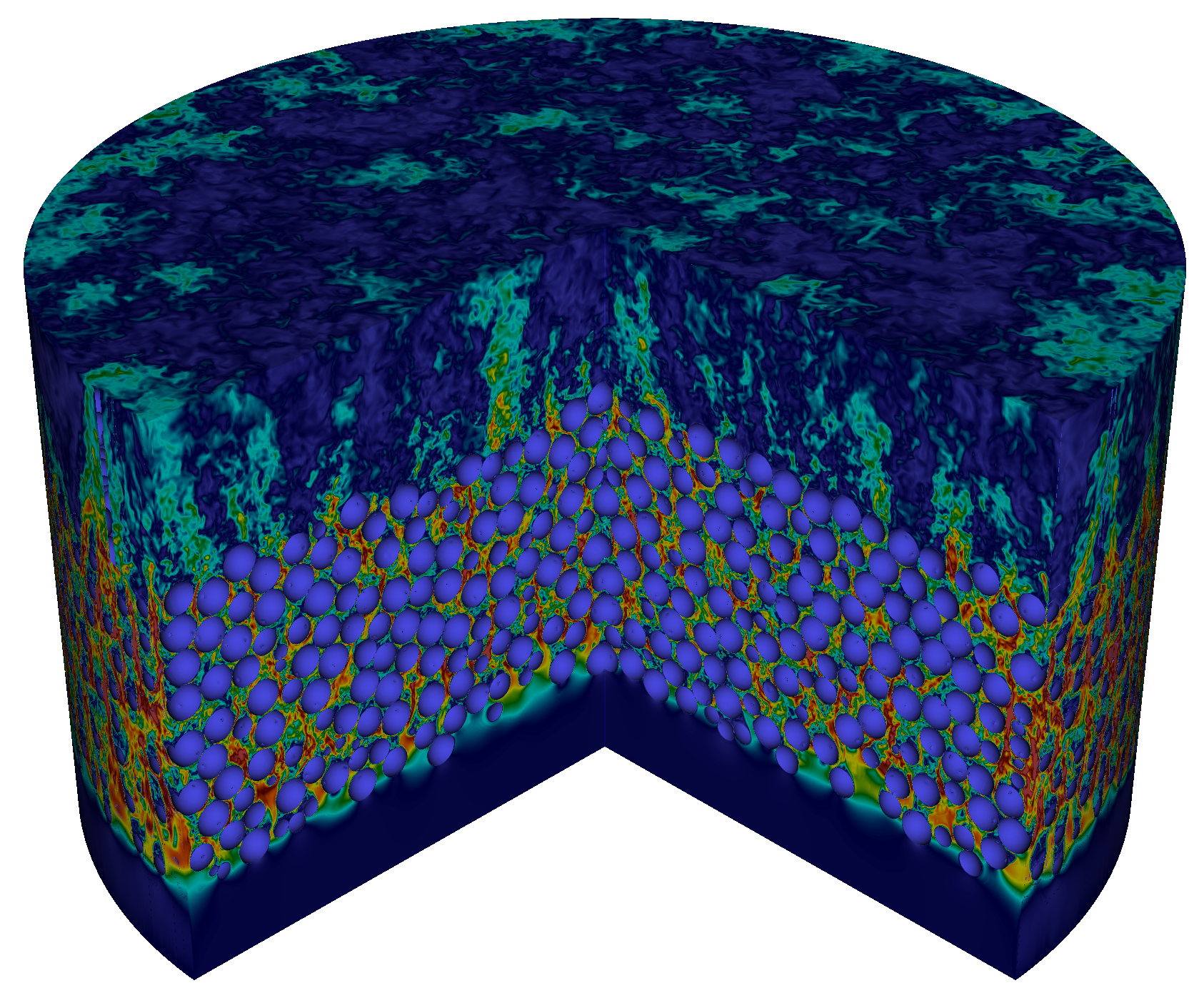}
    \end{subfigure}
    \begin{subfigure}[]{0.48\textwidth}
       \includegraphics[width=1.0\textwidth]{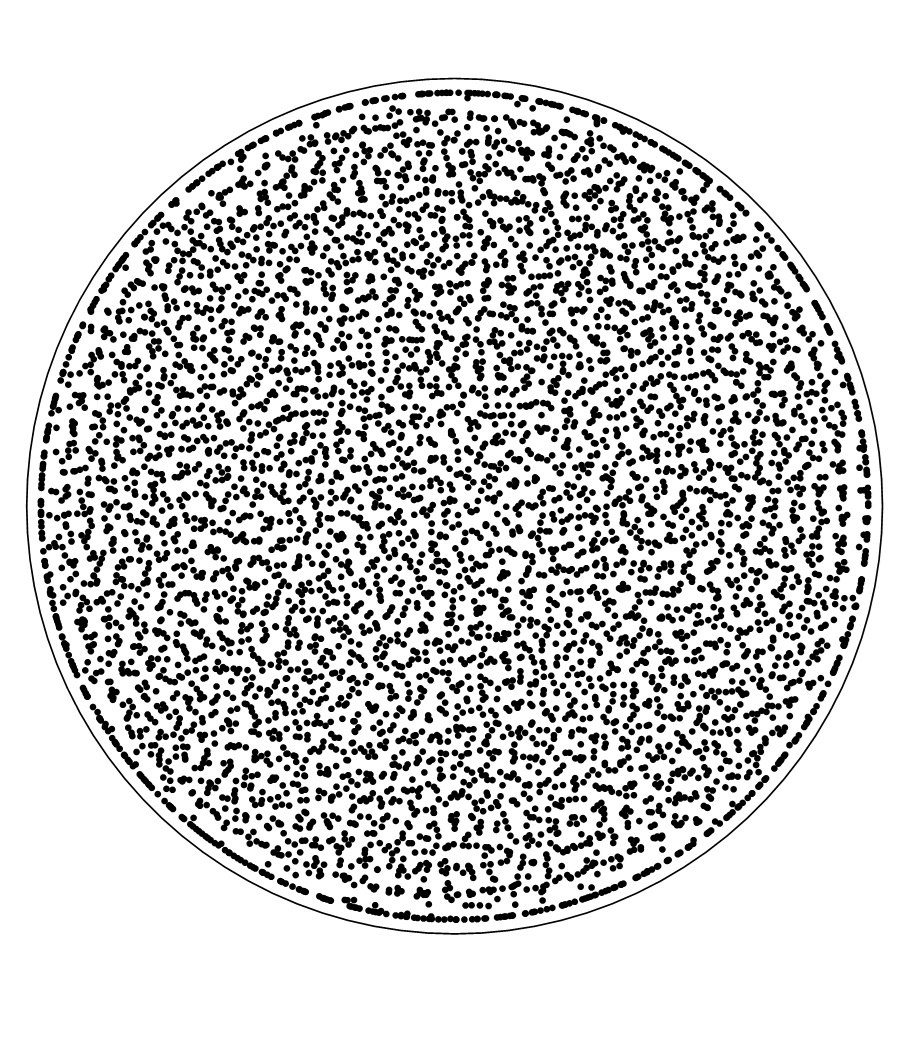}
    \end{subfigure}
    \caption{(left) Instantaneous velocity field for a bed of 7,000 pebbles ($D/d_{peb} = 30$). (right) Projection of the 7,000 pebble centers in a PBR onto an axial plane. The wall-channeling effect is visible in the organized ring of pebbles near the wall.}
    \label{fig:centers}
\end{figure}

One area where current closure models are insufficient is in the near-wall region of PBRs. In this region, the presence of the wall influences the packing of the pebbles, causing them to form more orderly structures and increasing the porosity. This effect can be seen in the projection of the pebble centers in a PBR found in Figure \ref{fig:centers}. Many existing closure models have been developed to predict the average behavior of the pebble bed as a whole, leading to inaccuracies when applied to the near-wall region that differs greatly from the bed interior. Accurate prediction of the flow in the near-wall region is critical, as it can have a significant impact on bypass flow where coolant is diverted through gaps between reflector blocks. Additionally, depending on the specific PBR design, there may be a peak in the fuel power near the wall where neutrons are reflected and thermalized by the graphite reflector. \textbf{Accurate modeling of the near-wall region is therefore critical to ensure accurate prediction of fuel temperature maximums and provide confidence in predicted safety margins.}  For these reasons, improving the understanding and modeling capabilities of the near-wall region has been identified by the United States Nuclear Regulatory Commission (NRC) as an issue of high importance \cite{PIRT}.

The near-wall region of pebble beds has been a topic of research interest for many years, with many researchers looking to improve understanding of this region to enhance modeling of the near-wall porosity, pressure drop, and heat transfer. The phenomena has been studied experimentally by Amini \cite{Amini2014}. Their study used hot wire anemometry probes to measure the near-wall flow velocities in several differently-shaped near-wall gaps. Their results experimentally confirmed the existence of the high-velocity flow channels that form in the gaps near the wall. They then examined the different behaviors observed in the two different near-wall gap geometries. Nguyen has also experimentally studied the flow behavior in the wall region through the use of Particle Image Velocimetry \cite{Nguyen2019}. They examined the cross flows between pebbles and the bypass flow in the near-wall region through the use of Proper Orthogonal Decomposition (POD). Computationally, the near-wall region has been studied by Fick and Merzari \cite{Fick2020}. Their study employed a regular arrangement of pebbles with a confining wall. They performed a Direct Numerical Simulation (DNS) and examined the 2nd and 3rd order flow statistics near the wall. Their study began to reveal some of the defining flow characteristics present in the near-wall region.
Modeling of the near-wall region has been improved through the use of porosity, pressure drop, and heat transfer equations. De Klerk developed a porosity correlation that is capable of accurately modeling the oscillatory porosity variation near the wall \cite{DeKlerk}. Reichelt was one of the first to investigate the effects of wall-channeling on the bed pressure drop. He found significant errors in the Ergun equation, one of the widely used pressure drop equations at the time, when applied to slender beds where the near-wall effects make up a large portion of the bed \cite{Reichelt1978}. He then suggested a new correlation that accounts for the ratio between pebble diameter and bed diameter. Eisefeld and Schnitzlein \cite{Eisfeld2001} identified the effects that wall-channeling may have on the pressure drop by comparing experimental results to correlations available at the time. They identified the approach by Reichelt as being promising in effectively modeling the influence of the near-wall region and developed their own improved correlation based on the findings from their study. With regards to heat transfer, work by Achenbach has studied the near-wall region and developed a correlation to model the near-wall heat transfer coefficient in packed beds \cite{Achenbach1986}.

This work reviews the current findings by our group at Penn State with regards to near-wall flow behavior in a pebble bed reactor \cite{RegerNT2022,RegerNED2023}. Additional investigation into the local geometry and Turbulent Kinetic Energy (TKE) production is then presented to better explain the observations that have been made thus far. Section \ref{sec:methods} details the methods used to study the near-wall flow, and results and analysis are then presented in Section \ref{sec:results}.

\section{Codes and Methods}
\label{sec:methods}
\subsection{Introduction to NekRS}
Argonne National Laboratory's spectral-element CFD code NekRS \cite{Fischer} was chosen as the high-fidelity code for this study. NekRS is a GPU-oriented variant of the well-established open-source code Nek5000 \cite{fischer2016}. It demonstrates excellent scalability \cite{merzari2020}, and is capable of linking to Nek5000 to utilize its existing pre- and post-processing features.

The simulations performed in this work use the incompressible, constant-properties Navier-Stokes equations in dimensionless form:

\begin{equation}
    \frac{\partial{\vec{v_i}}}{\partial{t}}+{\vec{v_i}}\cdot\nabla{\vec{v_i}}=-\nabla{P}+\frac{1}{Re}\nabla^2{\vec{v_i}}
\end{equation}
\begin{equation}
    \nabla\cdot{\vec{v_i}}=0
\end{equation}

where $v$ is the fluid velocity, $P$ is the fluid pressure, and $Re$ is the Reynolds number based on the pebble diameter and inlet velocity ($\frac{\rho{v_{inlet}}{D_{peb}}}{\mu}$). The nondimensionalization of variables follows the following scheme:

\begin{equation}
    x^*=\frac{x}{D_{peb}}
\end{equation}
\begin{equation}
    v^*=\frac{v}{v_{inlet}}
\end{equation}
\begin{equation}
    t^*=\frac{{t}{v_{inlet}}}{D_{peb}}
\end{equation}
\begin{equation}
    P^*=\frac{P}{\rho{v_{inlet}}^2}
\end{equation}

where $D_{peb}$ is the pebble diameter, $v_{inlet}$ is the inlet velocity, $t$ is the time, and $\rho$ is the fluid density. The simulations performed in NekRS are wall-resolved LES, where an explicit filter was used to approximate the effects of dissipation on the subgrid scales \cite{2021nek5000}. Simulations in NekRS were run with a polynomial order of 7.

\subsection{Mesh Creation}
The pebble beds created for high-fidelity simulation in this work were generated using the Discrete Element Method (DEM) in the open-source code Project Chrono \cite{projectchrono}.

The DEM simulations used to generate the beds for this work used the material properties of graphite, found in Table \ref{tab:properties} \cite{INLGraphite,XiaweiFriction,QiRestitution}. Additional information about the contact model used in Project Chrono can be found in \cite{chronocontact}. 

\begin{table}[]
    \centering
    \caption{Graphite material properties used in the DEM simulation. Properties were used for both sphere-sphere and sphere-wall contacts. \cite{INLGraphite,XiaweiFriction,QiRestitution}}
    \begin{tabular}{|c|c|}
        \hline
        Property & Value  \\
        \hline
         Density & 2260 $\mathrm{kg/m^3}$ \\
        Elastic Modulus & 8 GPa \\
        Poisson's Ratio & 0.12 \\
        Coefficient of Restitution & 0.6 \\ %DOUBLE CHECK THESE VALUES IN CASE I CHANGE THEM AT SOME POINT
        Sliding Friction Coefficient & 0.3 \\
        Rolling Friction Coefficient & 0.1 \\
        Simulation Timestep & $5\times10^{-5}$ s \\
        \hline
    \end{tabular}
    
    \label{tab:properties}
\end{table}

The beds were generated by randomly sampling sheets of pebbles separated by $2D_{peb}$ of axial distance. These sheets were then dropped into a cylindrical vessel to randomly pack. Several thousand pebbles were used for the packing, and then a section of pebbles was extracted from the center of the resulting bed to avoid any influence of the cylinder bottom or top. 

\begin{figure}
    \centering
    \includegraphics[width=0.5\textwidth]{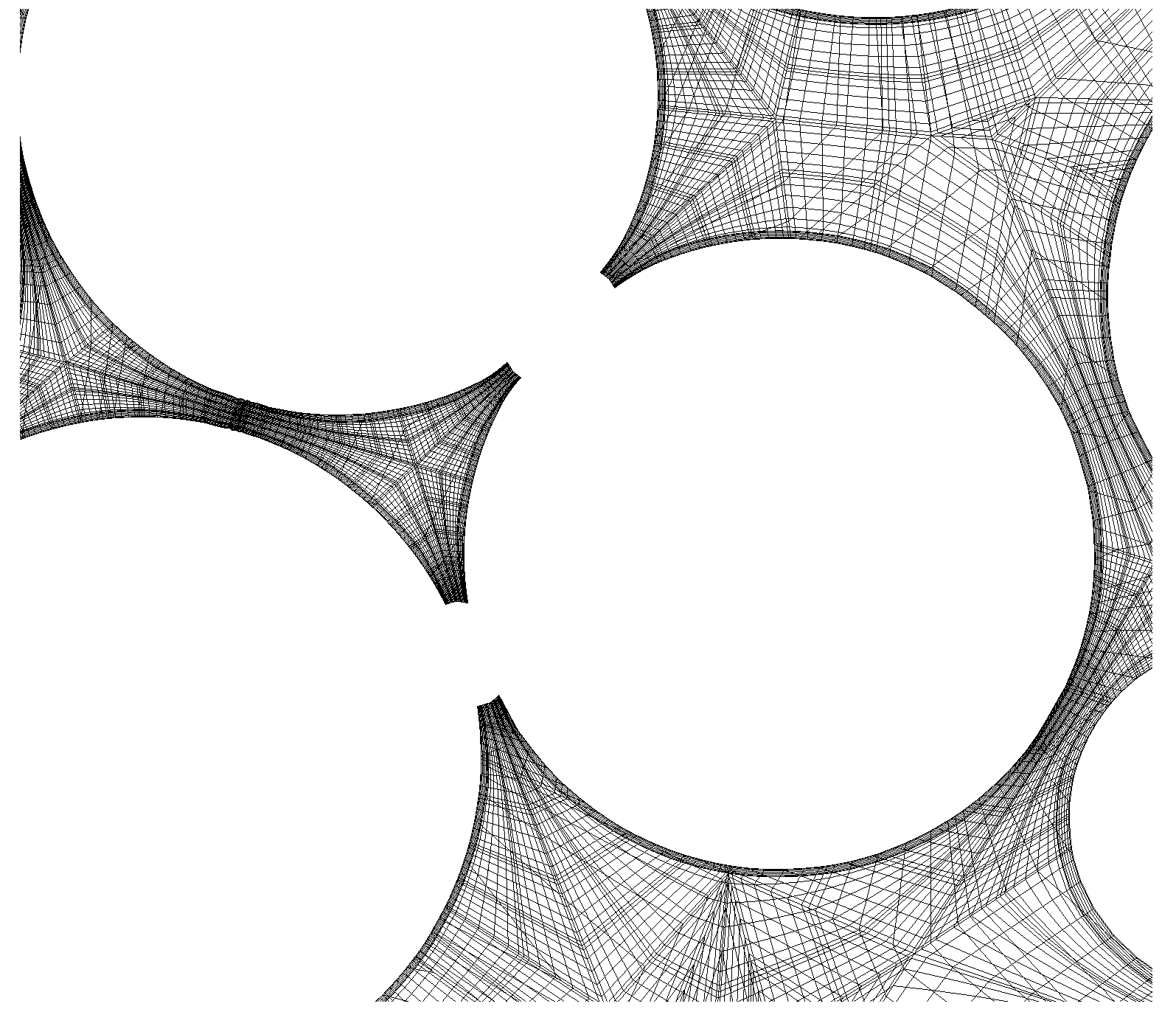}
    \caption{Example of the high-fidelity model meshes used for this work showing the chamfer between pebbles.}
    \label{fig:mesh}
\end{figure}

A meshing script is then used to generate an all-hexahedral mesh for simulation with NekRS. 
Developed as part of the Cardinal multiphysics project \cite{Cardinal}, the script uses a novel Voronoi-cell approach. It receives the pebble center coordinates along with the pebble and cylinder diameters as inputs and begins by generating a voronoi cell for the void region around each pebble. The faces of each voronoi cell are equidistant between pebbles. At the top and bottom of the bed, additional pebble centers are provided to determine the top and bottom voronoi faces, although these pebbles are not included in the resulting mesh. The voronoi faces are then modified to improve the resulting mesh quality by collapsing small edges or dividing very long edges. Quad elements are then generated on the faces and are projected down onto the pebble surfaces, producing an all-hexahedral mesh. Pebble-pebble and pebble-wall contacts are handled by adding a small chamfer at the point of contact. This simply inserts a small cylinder through the contact point to widen it slightly. This method has been shown to have a minimal effect on the resulting porosity and pressure drop compared to other methods, such as shrinking each pebble to avoid contacts \cite{LanMeshing}. An example of the resulting mesh is pictured in Figure \ref{fig:mesh}.

We note that we designed the mesh to have resolution sufficient to resolve the Taylor micro-scales based on estimates from a previous work \cite{Yildiz} and additional calculations computed for this work with RANS. We also design the mesh to resolve the boundary layer and have the first grid point at $y^{+}<1$ and a sufficient number of points in the viscous sub-layer. 

A flat-profile inlet velocity condition and a stabilized outflow condition \cite{dong2014} are then applied along with no-slip conditions on the pebble and cylinder walls.

\subsection{Turbulent Kinetic Energy Budgets}
The Turbulent Kinetic Energy (TKE) represents the mean kinetic energy that is carried by the eddies found in turbulent flow. TKE is typically produced through shear, forcing, or friction. It is then transferred donw the energy cascade where it is eventually dissipated by viscous forces in the smallest eddies. The TKE can be represented as the sum of the individual contributing mechanisms to produce the TKE equation:

\begin{equation}
    \underbrace{\frac{\partial{k}}{\partial{t}}}_\text{TKE Derivative}+\underbrace{\overline{u_j}\frac{\partial{k}}{\partial{x_j}}}_\text{Advection}=\underbrace{-\frac{1}{\rho}\frac{\partial\overline{u'_{i}{p'}}}{1}}_\text{Pressure Diffusion}-\underbrace{\frac{1}{2}\frac{\partial\overline{{u'_j}{u'_j}{u'_i}}}{\partial{x_i}}}_\text{Turbulent Transport}+\underbrace{\nu\frac{\partial^2{k}}{\partial{x}^{2}_{j}}}_\text{Molecular Transport}-\underbrace{\overline{{u'_i}{u'_j}}\frac{\partial\overline{u_i}}{\partial{x_j}}}_\text{Production}-\underbrace{\nu\overline{\frac{\partial{u'_i}}{\partial{x_j}}\frac{\partial{u'_i}}{\partial{x_j}}}}_\text{Dissipation}
\end{equation}

These terms are commonly referred to as the budgets of the TKE. Analysis of the production term for a packed bed is carried out in section \ref{sec:results}.

\subsection{Validation of NekRS}
\label{sec:val}
\begin{figure}
    \centering
    \includegraphics[width=0.9\textwidth]{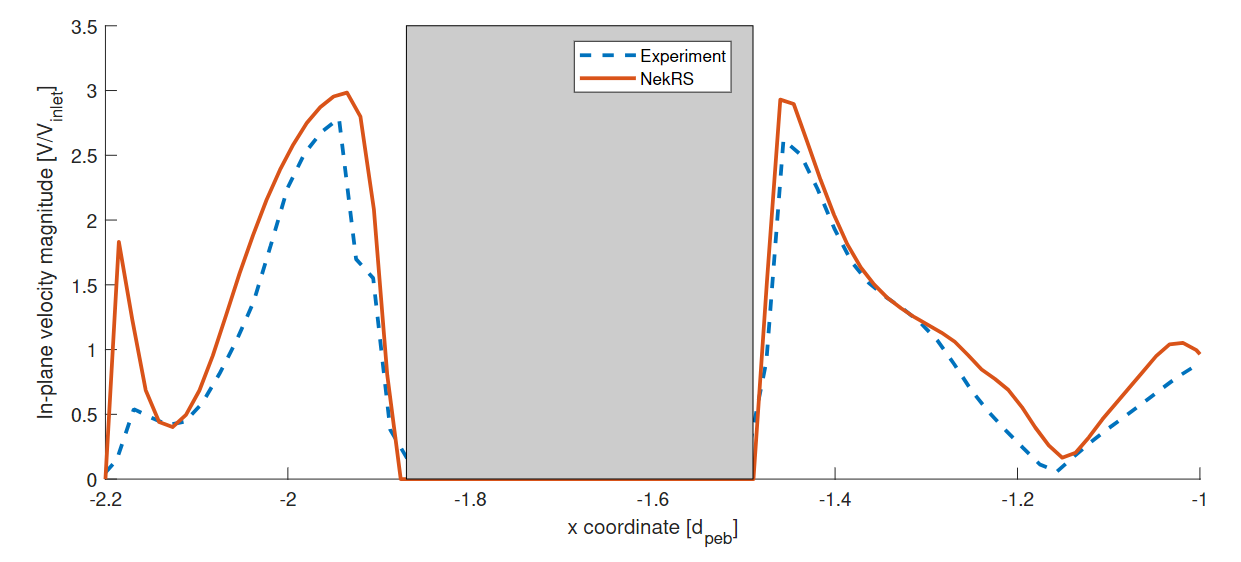}
    \caption{In-Plane velocity magnitude over one sampling line for the 67-pebble validation study. The shaded gray region indicates the location of a pebble. Retrieved from \cite{Reger_NT2023}}
    \label{fig:velval}
\end{figure}

\begin{figure}
    \centering
    \includegraphics[width=0.7\textwidth]{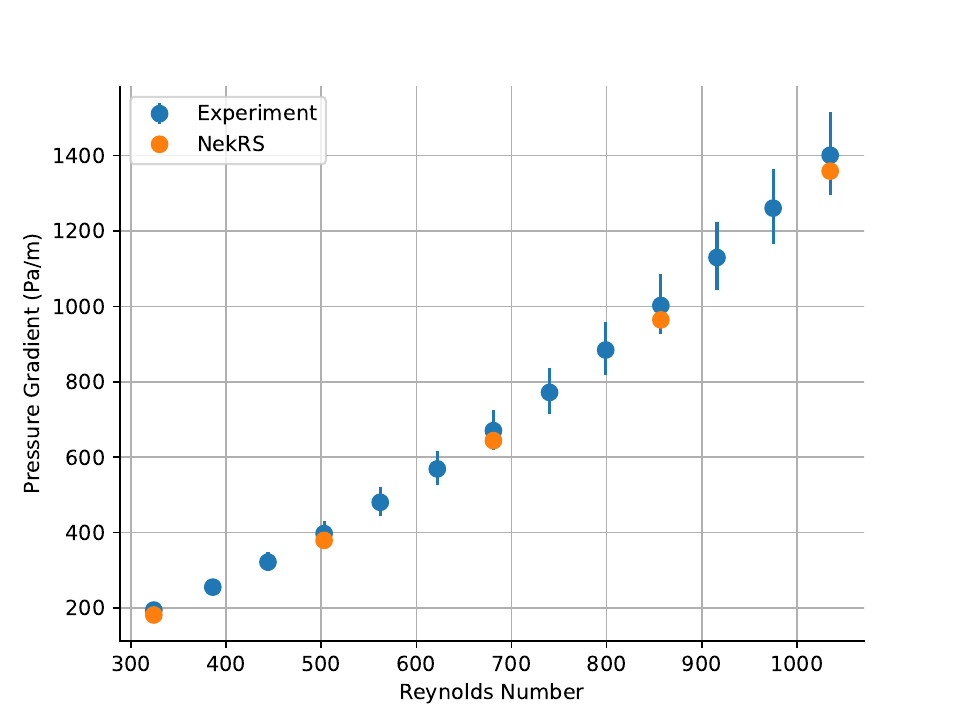}
    \caption{Comparison of pressure gradients between experiment and corresponding simulation.}
    \label{fig:presval}
\end{figure}

Validation of NekRS's ability to reproduce the velocity and pressure drop in a packed bed has been performed with experimental data from Texas A\&M University. Velocity validation was performed on a bed of 146 pebbles in work by Yildiz \cite{Yildiz}. Additional validation has been performed for a bed of 67 pebbles, a comparison of velocity profiles between experiment and simulation can be seen in Figure \ref{fig:velval}. Validation of the pressure drop was performed by comparing results between experimental and simulation pressure gradients over five Reynolds numbers for a second bed of 789 pebbles. This comparison is shown in Figure \ref{fig:presval}, where it can be seen that NekRS falls within the range of experimental uncertainty at all points of comparison.

\section{Results}
\label{sec:results}
\subsection{Near-Wall Form Coefficient Trends}
\begin{figure}
    \centering
    \includegraphics[width=0.7\textwidth]{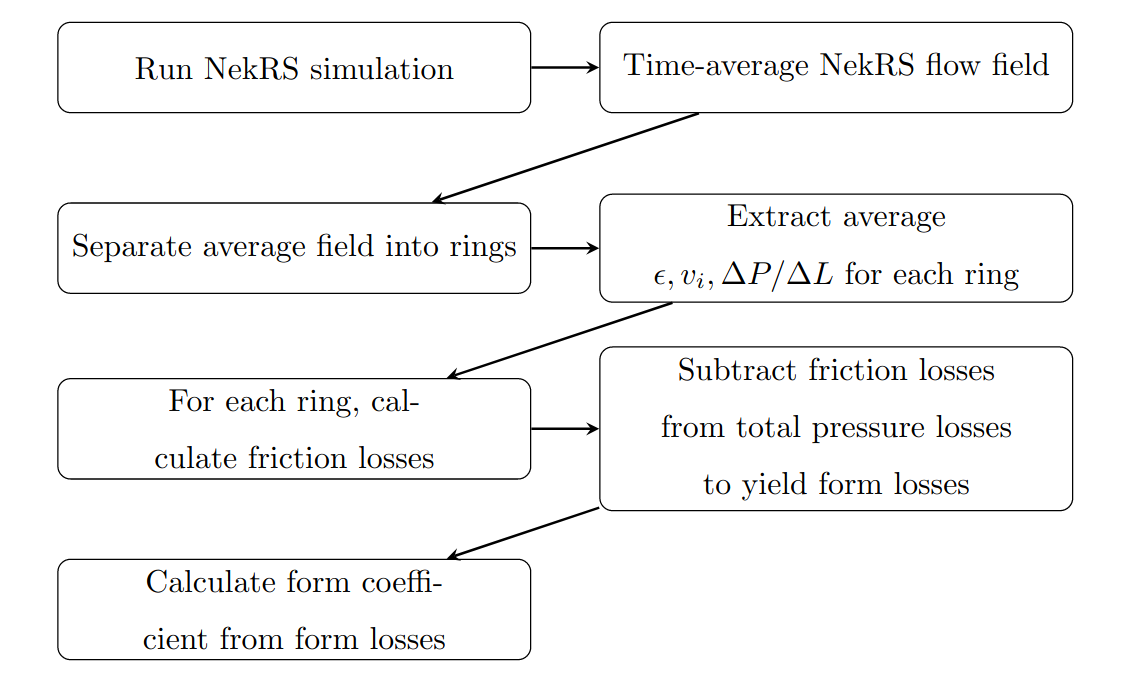}
    \caption{Data extraction method used to generate the improved pressure drop equation from \cite{RegerNED2023}.}
    \label{fig:method}
\end{figure}

Current work by our group has investigated the effect of the cylinder wall on the form loss coefficients \cite{RegerNT2022,RegerNED2023}. An understanding of this effect is critical to properly model the radial variation in the streamwise velocity of a PBR. The goal of this study was to improve the capabilities of the KTA drag correlation\cite{KTA} to more accurately model the near-wall velocity profile in a porous media model. NekRS simulation was used to produce an LES flow dataset for two pebble beds of roughly 1,568 and 1,700 pebbles at aspect ratios ($D_{bed}/D_{peb}$) of 13 and 14. This analysis was then performed by separating the beds into concentric rings of $0.05D_{peb}$ width. The ring-volume-average porosity, fluid velocity, and pressure drop could then be calculated for each ring. This information can then be used to calculated the form loss coefficient in each ring. An overview of the data extraction method is shown in Figure \ref{fig:method} with more information available in Reference \cite{RegerNED2023}.

\begin{figure}
    \centering
    \includegraphics[width=0.7\textwidth]{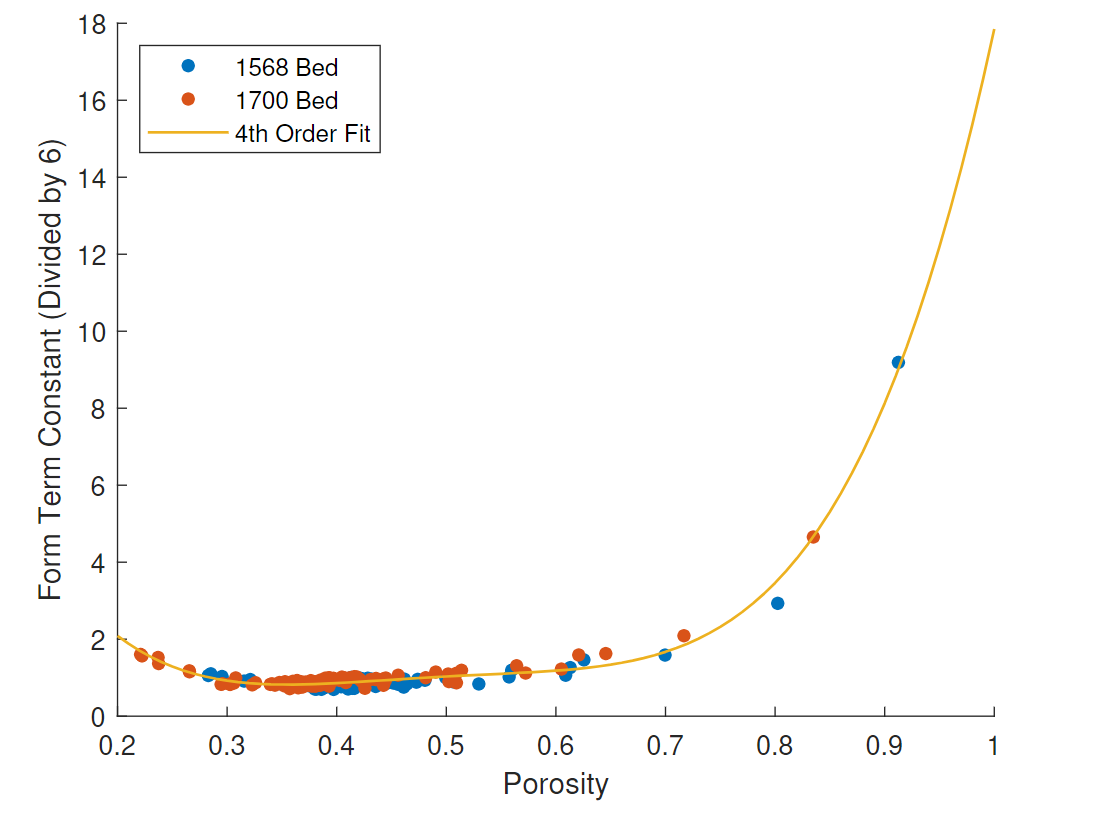}
    \caption{Form loss term constants plotted against the porosity of each respective ring. A fourth order fit is applied to the data.}
    \label{fig:formconst}
\end{figure}

The form loss terms were then plotted against the Reynolds number in each ring, and the constant in the numerator of the KTA form term was calculated for each ring. These constants were then plotted against the ring porosity, shown in Figure \ref{fig:formconst}. It was determined that there is a correlation between the form constant and the ring porosity. A fourth-order polynomial was fit to the data to describe this correlation:
\begin{equation}
    f(\epsilon) = 253.9\epsilon^4-499.3\epsilon^3+364.7\epsilon^2-115.6\epsilon+14.21
\end{equation}

where $\epsilon$ is the local ring porosity.
This correction term may be added to the KTA equation to produce an improved drag correlation:
\begin{equation}
    \frac{\Delta{P}}{L} = \left(\frac{320}{{Re}_{m}}+\frac{6{f(\epsilon)}}{{{Re}_{m}}^{0.1}}\right)\left(\frac{1-\epsilon}{\epsilon^3}\right)\left(\frac{\rho^2{v_{s}}^2}{{D}_p}\right)\left(\frac{1}{2\rho}\right)
    \label{eq::RMKTA}
\end{equation}

It should be noted that this correlation can currently be stated to be valid for $10^2<Re_m<10^4$ and $0.2<\epsilon<0.9$. Additional data may help to further quantify the uncertainties of this correlation at the high and low limits of the porosity where the data is currently sparse.

After the form analysis was performed and a new correlation was determined, several knowledge gaps still persisted. Most importantly, it was not immediately clear as to \textit{why} this correlation between the region porosity and the form loss existed. Similarly, it was also unclear as to whether this variation is truly a correlation with porosity, or rather with wall distance. The next section presents some preliminary analysis into possible explanations for the dramatic increase in the form loss constant as the ring porosity increases.

\subsection{Toward an improved understanding of the wall-channeling effect}

\begin{figure}
    \centering
    \includegraphics[width=0.7\textwidth]{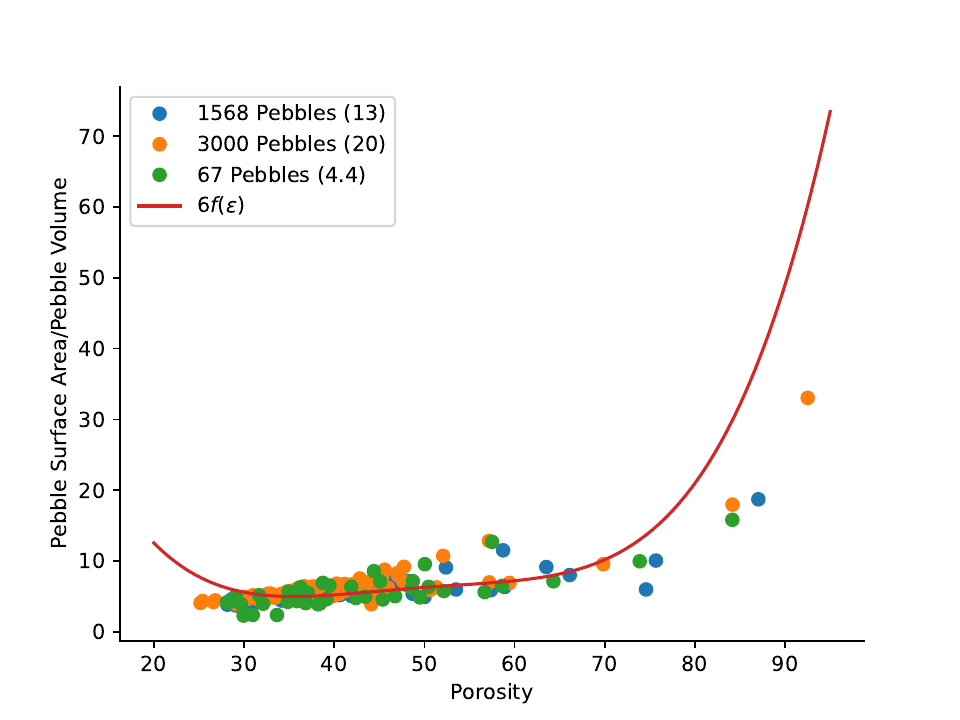}
    \caption{Ratio between pebble surface area and pebble volume $S_v$ calculated for rings of $0.05D_{peb}$ width versus ring porosity for several pebble beds. The form constant term with correction factor, $6f(\epsilon)$ is also shown. All beds are nondimensionalized to $D_{peb}=1$ with the bed aspect ratio shown in parentheses in the legend.}
    \label{fig:sv}
\end{figure}

Among previously derived equations for the pressure loss in a packed bed of spheres, the pebble surface to volume ratio $S_v$ is commonly used. Carman and Kozeny \cite{Carman1937} used this definition in their derivation of the viscous and inertial losses in a packed bed. It was also used by Ergun \cite{Ergun} who built off of the work of Carman and Kozeny. In these previous derivations, $S_v$ of a sphere is used, which can easily be calculated as $6/D_p$. The actual surface to volume ratio was calculated for concentric rings of $0.05D_{peb}$ width for several computational pebble beds of different sizes. These values were then plotted against the porosity in each ring, shown if Figure \ref{fig:sv}. From this plot, it can be seen that $S_v$ is roughly equal to $6/D_{peb}$ in low and medium porosity regions. The average porosity of most beds falls around 0.35-0.5 depending on the bed aspect ratio, meaning that the $6/{D_p}$ value that many correlations have assumed for $S_v$ is fairly accurate when applied to an averaged bed. As the porosity increases, however, this value increases significantly. Also shown in Figure \ref{fig:sv} is the trend of the form constant $6f(\epsilon)$ for comparison. The trend in $S_v$ is significantly more linear than that of $f(\epsilon)$, although this increase is the surface to volume ratio may be a contributor to the additional dependency of the form constant on the porosity. 

Pressure drop equations such as the Ergun, Carman-Kozeny, KTA, and our improved KTA equation separate the porosity effects from the viscous and inertial loss coefficients. This has previously allowed for the representation of the losses with a quantity that represents the losses per pore and a porosity-dependent quantity that represents the number of pores. The multiplication of these two terms then yields a drag coefficient. The result from Figure \ref{fig:sv} suggests that this previous approach may not be accurate, as the loss per pore changes as a function of the pore shape (which is itself a function of the porosity). This result suggests higher losses per pore as the porosity is increased. This at first may seem counter intuitive, as one would expect a more open and uniform flow geometry to have lower pressure loss. This intuition remains correct even with the findings from Figure \ref{fig:sv}, as although the losses per pore are suggested to be higher at high porosities, the total loss coefficients will still remain lower than a lower-porosity region, as the $\frac{1-\epsilon}{\epsilon^3}$ term that represents the number of pores will be small for high porosity regions.

\subsubsection{Investigation of the Turbulent Kinetic Energy production}
\begin{figure}
    \centering
    \begin{subfigure}[]{0.48\textwidth}
        \includegraphics[width=\textwidth]{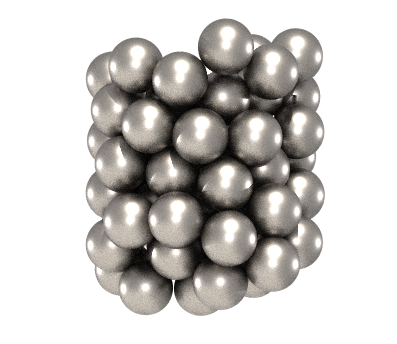}
    \end{subfigure}
    \begin{subfigure}[]{0.48\textwidth}
        \includegraphics[width=\textwidth]{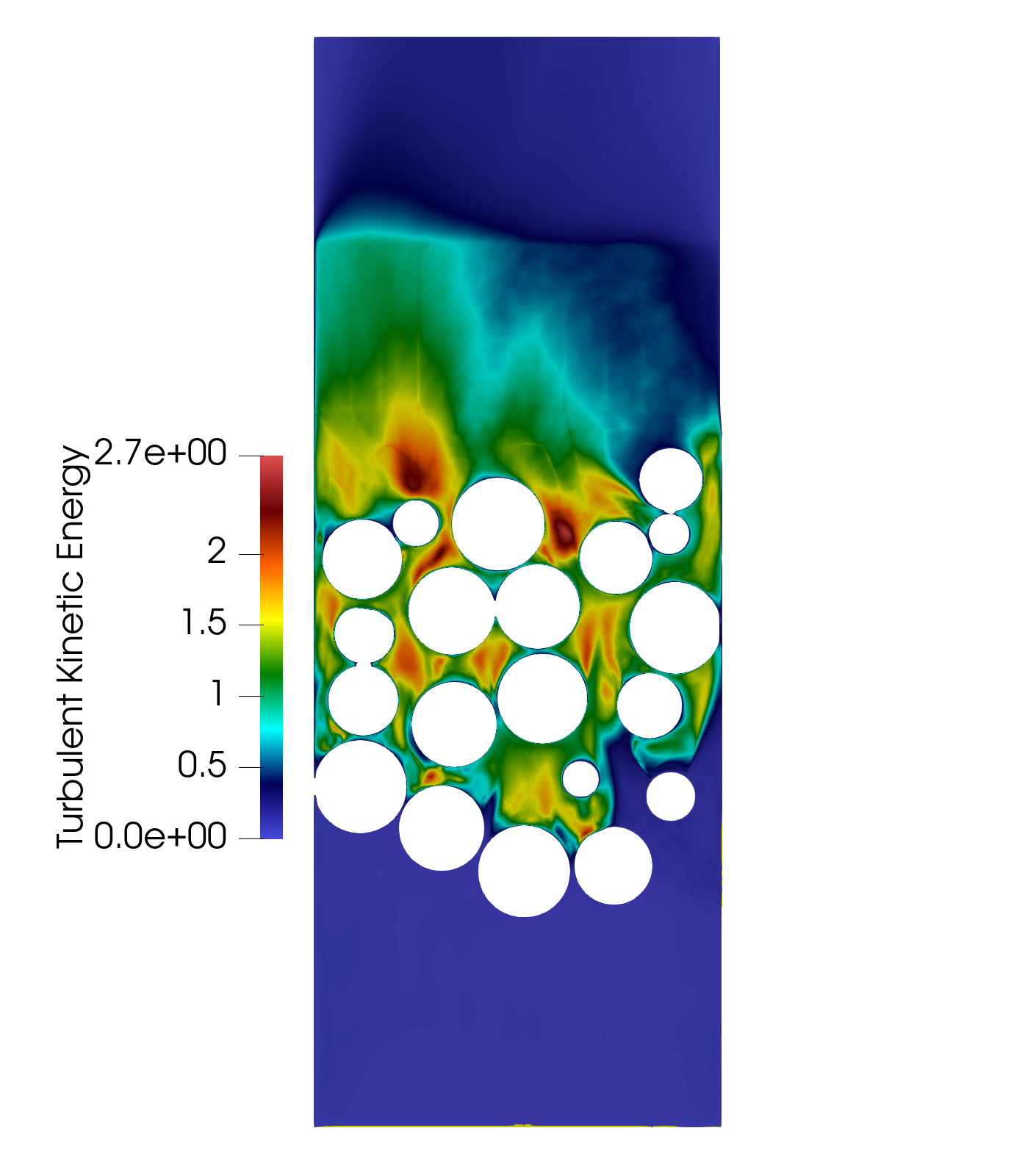}
    \end{subfigure}
    \caption{67 pebble bed used for the DNS simulation (left) and centerplane slice of the TKE (right).}
    \label{fig:tke}
\end{figure}

A better understanding of the near-wall flow physics can be obtained by examining the budgets of the TKE in the near-wall region. The budget terms were calculated from a Direct Numerical Simulation (DNS) of a small bed of 67 pebbles at Re = 1,460 that was previously used for experimental velocity validation in section \ref{sec:val}. Previous work examined line samples of the fluid velocity at three locations \cite{Reger_NT2023}, one of which is shown in Figure \ref{fig:velval}. For the DNS in this work, the bed was simulated at a polynomial order of 9 for 50 convective units to converge the 3rd-order statistics. The bed is shown alongside a centerplane slice of the TKE in Figure \ref{fig:tke}. Some of the notable features are the peaks in the TKE, typically in the wake regions behind the pebbles. There is also a noticeable left-side bias in the TKE at the bed outlet. This is can be attributed to the distribution of the top layer of pebbles, causing a turbulent plume in this region. Analysis of the TKE production term, as was done in previous work \cite{Fick2020}, can reveal additional information on the inertial effects throughout the bed. 

\begin{figure}
    \centering
    \includegraphics[width=0.7\textwidth]{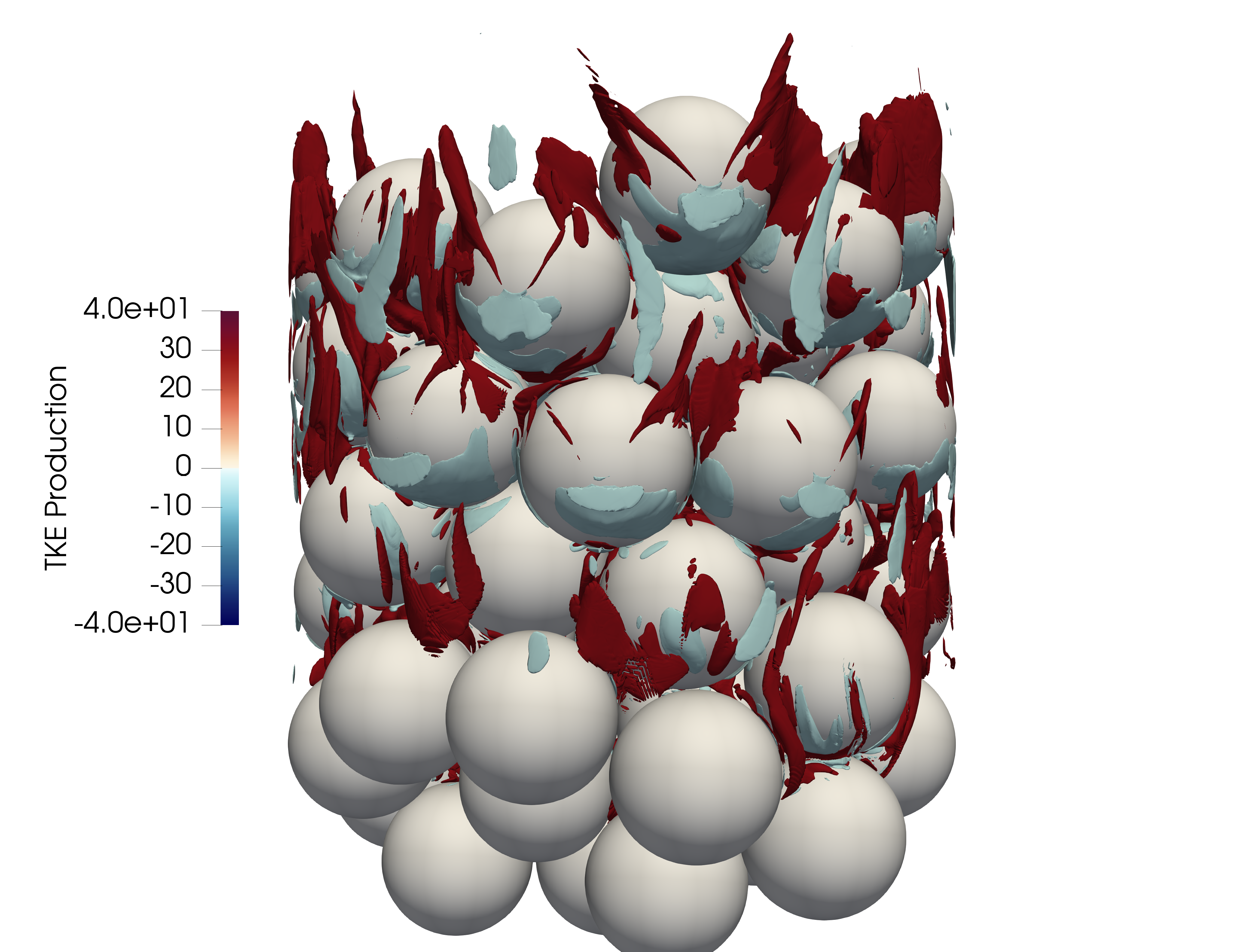}
    \caption{3D Visualization of isosurfaces enclosing areas where the TKE production is less than or equal to -5 and greater than or equal to 30.}
    \label{fig:prd}
\end{figure}

Figure \ref{fig:prd} shows the isosurfaces of highly positive and negative TKE production in the bed. The areas of negative production are particularly interesting. In many canonical cases, such as channel flow, the production term is positive and acts as a source of TKE. A complex case such as a pebble bed, however, can see this term change sign and become negative, suppressing the TKE rather than strengthening it. There are large areas of negative production on the bottom surfaces of the pebbles. Flow accelerates around the pebble in these regions, causing negative production and suppression of the TKE. Additionally, there are some elongated negative-production structures in the large void regions near the wall. The areas of high TKE production exist in the wake regions behind pebbles where the velocity gradients and covariances are high.

\begin{figure}
    \centering
    \includegraphics[width=0.7\textwidth]{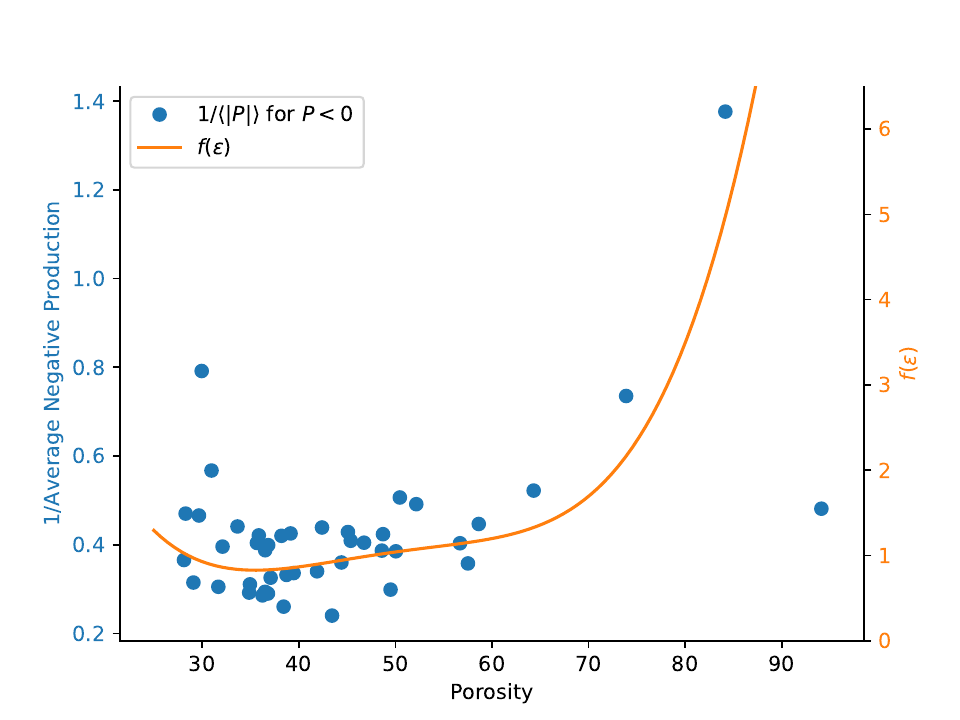}
    \caption{Average of negative production ($1/\langle|P|\rangle$ for $P<0$)  versus local ring porosity. The correction factor $f(\epsilon)$ is also included on a separate axis for reference.}
    \label{fig:negativeProd}
\end{figure}

Further investigation into the TKE production was also performed to determine if there is any relationship to the observed form correction factor. It was theorized that differences in the amount of negative production could potentially explain the correlation between the form constant and the porosity that was observed previously. As has been done with the other investigations, the 67-pebble bed was separated into rings of $0.05D_{peb}$ width, and the average negative TKE production ($\langle|P|\rangle$ for $ P<0$) was calculated for each ring. An inverse relationship between the negative TKE production and the the porosity was found that closely matches the relationship between the form constant and the porosity. Figure \ref{fig:negativeProd} presents this relationship, along with the correction factor $f(\epsilon)$ for reference. It can be seen that the trends of $1/\langle|P|\rangle>$ with the porosity exhibit many similarities to the trends of $f(\epsilon)$, as there are increases in both the high and lower porosity regions with a slight linear increase in the medium-porosity regions. This points to the idea that inertial flow effects vary greatly based on the porosity as a result of the different void geometries. These inertial effects will have an effect on the form loss coefficient which can perhaps explain the trends described with the $f(\epsilon)$ correction factor. Regions with high negative TKE production experience more laminarization and a lower form loss per pore. Meanwhile, areas with lower negative TKE production exhibit the opposite behavior, with less laminarization and higher form loss per pore, aligning with the observed trend in the form constant found in Figure \ref{fig:formconst}. The data point nearest to the wall in Figure \ref{fig:negativeProd}, however, still requires additional investigation, as it does not follow the trend seen with the rest of the data. The bed used to extract this data is fairly small, leading to small averaging volumes and poor averaging statistics. Although the results presented from the investigation of the negative production suggest a possible explanation for the trend in the form constant, it remains difficult to decisively confirm this explanation. Additional data for larger beds to gather additional data points is necessary to further reinforce this claim.

\section{Conclusion}
This study presented a review of current efforts at Penn State University to gain a better understanding of the near-wall region of pebble bed reactors. Work has thus far studied the flow in a PBR by dividing the bed into multiple concentric regions and examining temporally and spatially-averaged flow characteristics in each region. Examination of the form-loss coefficients has revealed a correlation between the form loss term and the porosity of each ring region. This information has been used to generate an improved pressure drop correlation that is capable of more accurately reproducing the radial velocity profile in a PBR with a porous media code \cite{RegerNED2023}.

Although this improved correlation has proven to be a promising result, additional work is currently being conducted to better understand the near-wall flow phenomena that may be behind the observed correlation between the form term and the porosity. This study investigated the geometry of the concentric rings, particularly the ratio between the solid surface area and the solid volume. This ratio (denoted $S_v$) has been used as part of the derivation of many past pressure drop equations, where it has been assumed to be the theoretical value of $6/D_p$. Calculation of $S_v$ for several computational beds revealed that $S_v$ actually varies with the porosity, ranging from $3/D_p$ to greater than $30/D_p$ based on the porosity. This helps to explain the increase in the form constant term with the porosity, although the relationship between $S_v$ and $\epsilon$ is fairly linear. This indicates that although $S_v$ may play a role in influencing the increasing form constant, it is likely not the only contributor.

The TKE production in each ring was then investigated to further examine potential flow phenomena that may contribute to the varying form constant. A relationship was found between the inverse of the negative TKE production and the porosity that closely matches the trend observed in the form constant. This suggests that the observed trend in the form constant with the porosity may be related to inertial effects that are caused by the different pore geometries at different porosities. The bed used to generate the TKE budgets was rather small, at only 67 pebbles, and thus further data for larger beds is needed to reinforce the hypothesis on the effect of the negative production.

\section*{ACKNOWLEDGMENTS}

This material is based upon work supported under an Integrated University Program Graduate Fellowship.

This research used resources of the Oak Ridge Leadership Computing Facility at the Oak Ridge National Laboratory, which is supported by the Office of Science of the U.S. Department of Energy under Contract No. DE-AC05-00OR22725.

This research made use of Idaho National Laboratory computing resources which are supported by the Office of Nuclear Energy of the U.S. Department of Energy and the Nuclear Science User Facilities under Contract No. DE-AC07-05ID14517.

%Bibliography
\bibliographystyle{unsrt}  
\bibliography{references}

\end{document}